**Title:** Do the rich pay their fair share? Enumerating the financial and emissions consequences of abolishing premium air travel

**Authors:** Megan Yeo[1], Sebastian Nosenzo[2], Daniel S. Palmer[3], Alexei K. Varah[4], Lucas Woodley[5], and Ashley Nunes[1,6]

[1] Department of Economics
Harvard College
Cambridge, MA, 02138, USA

[2] Graded - The American School of São Paulo
São Paulo - SP, 05642-001, Brazil

[3] Department of Economics
University of Chicago
Chicago, IL, 60637, USA

[4] Hunter College High School
New York, NY, 10128, USA

[5] Graduate School of Arts and Sciences
Harvard University
Cambridge, MA, 02138, USA

[6] Center for Labor and a Just Economy
Harvard Law School
Cambridge, MA, 02138, USA

**Corresponding Author:** Ashley Nunes, anunes@fas.harvard.edu

**Abstract**

Premium air travel – driven by demand from upper income households – faces increasing scrutiny given its disproportionately large carbon footprint. This footprint, existing discourse suggests, is driven by the increased space and amenities typically found in premium cabins. One increasingly popular solution to abate these emissions is to disincentivize or prohibit premium air travel in its entirety, favoring instead all-economy cabins. What might the impact of such a policy be? We address this question by leveraging an empirical model that integrates cabin configuration data, fuel burn profiles across various aircraft types, and airfares for over 12,000 flights. Our findings are threefold. First, we find that removing premium cabins in their entirety can reduce per-passenger emissions by between 8.1 percent and 21.5 percent, depending on the type of aircraft and aircraft stage length. Second, we observe that on a per-flight and lifespan basis, removing premium cabins offers far fewer, if any, reductions in emissions. Our estimates suggest that all-economy flights can reduce emissions by 0.45 percent or increase emissions by up to 1.43 percent on a life-cycle basis. Third, we ascertain that all-economy configurations reduce aggregate airline revenue by between 4.92 percent and 23.1 percent. Given these reductions, maintaining baseline revenue requires a 6 percent to 30 percent increase in airfare ticket prices, which levies a greater cost burden on working-class travelers. We conclude that foregoing premium cabins may fail to adequately reduce aggregate emissions while hindering air travel accessibility via higher airfares.

**Overview**

Addressing climate change requires rapid sector-wide decarbonization (1). The aviation sector is a key target of such decarbonization efforts, as flying is one of the most carbon-intensive modes of travel (2,3). However, aviation is a particularly challenging sector to decarbonize due to its dependence on fossil fuels with high energy density, such as kerosene, that provide the necessary thrust to overcome gravity (4)[1]. Burning kerosene imposes significant climate externalities, the most notable being the release of greenhouse gases, especially carbon dioxide.

Given the potential impact of these externalities, efforts to temper carbon emissions from the aviation sector have been far-ranging (6-15). These include leveraging electric (rather than kerosene) propulsion systems, using synthetic aviation fuels, and achieving more efficient kerosene consumption through operational improvements (6-9). Challenges persist for each of these pathways. Low energy density (compared to kerosene) currently characterizes electric batteries, which invariably limits the range of battery-powered aircraft, and, by consequence, what routes these aircraft can fly (6). Synthetic aviation fuels remain between 120 percent and 700 percent more expensive than fossil fuels, in part due to a lack of adequate supply (7,8). Finally, reduced kerosene consumption, achieved through operational efficiencies (e.g., by utilizing more fuel-efficient aircraft and optimizing airspace structure), delivers a fraction of the emissions reductions required to meet current climate goals (e.g., the Paris Agreement) (8,9).

One potential solution to aviation's high carbon footprint that has been gaining increasing traction is disincentivizing the use of premium cabins, either via taxation or regulatory prohibition (10,11). Existing literature suggests that the 'carbon footprint' of these cabins (relative to economy class cabins) is significant, due to business and first-class seats displacing proportionately more economy seats for the same total aircraft space capacity (10,12). This displacement and the associated emissions consequence reflect the larger seat space afforded to premium flyers and the characteristics of premium seats (e.g., lie-flat functionality, ultra-widescreen TVs, and dedicated hangar space) that make them heavier and consequently, more energy-intensive to move than economy seats. Proponents argue that eliminating these seats (and consequently, premium travel classes) would facilitate emissions reductions by concentrating passengers on fewer flights (13-15).

How effective might such a proposition be? What are the emissions consequences of eliminating premium travel? Do foregoing premium travel classes produce measurable reductions in emissions? If so, how profound are these reductions? And what might the financial impact of obfuscating premium travel be, both from the vantage point of airlines and passengers? Our work addresses these questions. Leveraging publicly available data on aircraft seating capacity, load factors, and fuel requirements, we estimate a model that scrutinizes the emissions footprint of premium versus non-premium travel. Unlike previous work, we also acknowledge that emissions are likely to vary based on flight-specific factors, the most significant of which is flight duration (hereafter referred to as stage length), and aircraft type (10). Observing that stage length and aircraft type are often confounded with one another (i.e., larger aircraft boast longer ranges and are hence more likely to be used for routes characterized by longer stage lengths), our model allows for a decoupling of these parameters.

---

[1] During takeoff and climb, engines must generate large amounts of thrust to overcome gravity and reach cruising altitude. Consequently, while these stages are – in temporal terms – short, the fuel consumed (and by consequence, the associated emissions generated) are – in proportional terms – large (5).

Most notably, our efforts differ from previous work in that we consider the financial impact of eliminating premium travel classes. We do so from the vantage point of both airlines and passengers, leveraging publicly available airfare data to assess how all-economy flights may affect airlines' revenue and passengers' airfares. This approach reflects an appreciation for the capital-intensive nature of air travel, the slim margins endemic to the airline industry, and the crucial role that premium flyers play in ensuring airline solvency through revenue generation (13, 17). Premium flyers generate over 50 percent of an airline's revenue despite constituting only 12 percent of an airline's passengers (13). Consequently, accounting for the potential loss of this revenue source is timely.

The current trajectory of aviation emissions evidences the importance of our efforts. While emissions in this sector currently account for approximately 2.5 percent of global carbon emissions, this figure is expected to rise to 22 percent by 2050 due to projected increases in air travel demand. Compounding this challenge is the greater warming effect associated with aviation (compared to other sectors) owing to the high-altitude release of greenhouse gases that contribute to contrail formation, trapping the Earth's heat (18,19). Given the challenging nature of reducing emissions in the aviation sector, policies that facilitate emissions reductions – without impeding the economic benefits afforded by aviation – warrant scrutiny. Our efforts scrutinize the viability of such a policy.

**Method**

To assess the emissions and financial consequences of eliminating premium travel cabins, we estimate premium cabins' current emissions and financial impacts using data on three key inputs: 1) space/seat allocation on an aircraft across all classes of travel (e.g., economy, premium economy, and business); 2) emissions profile of the aircraft; and 3) revenue generated by passengers in each travel class. We note that emissions and revenue data are generated for both short-haul and long-haul travel, as doing so affords an added layer of granularity in our analysis. Given the lack of universal agreement on what constitutes a short-haul versus long-haul route, we define short-haul routes as covering approximately 631 nautical miles (with a roughly 2-hour block time) and long-haul routes as covering approximately 3,003 nautical miles (with a roughly 7-hour block time).

Below, we provide an overview of each data input, detailing how each was sourced and leveraged by our model. Given the impracticality of sourcing such data for all airlines, routes, and aircraft types, we focus on a subset that reflects global brand recognition, connectivity potential (i.e., the number of destinations served by the carrier), and passenger carriage volume (i.e., the number of passengers using the airline annually).

Cabin Space Allocation: Our model uses space/seat allocation data for four aircraft types: the Airbus A320-200 and Boeing 737-800, which typically serve short-haul markets, and the Airbus A330-200 and Boeing 777-200LR, which typically serve long-haul markets. For each aircraft type, we assess the precise cabin layout employed by four airlines: Air Canada, Delta Airlines, ITA Airways, and Turkish Airlines. This sample set yields 16 unique aircraft-airline space combinations. Assessment of cabin layout entails enumerating total available space in the passenger cabin, space allocated to each travel class in the cabin (i.e., business, premium economy, and economy), number of seats in each of those cabins, and space allocated for non-passenger seating purposes (e.g., galley, toilets, aisles, cockpit) (hereafter referred to as service zones) for each aircraft type and airline (see Supplementary Information: Table S1a). Data is averaged across all four airlines operating a specific aircraft type to create a composite cabin for each aircraft type (yielding a total of four composite cabins) (see Table 1 and Figure 1). For example, the current seating capacities for the Airbus A320-200 are 146, 157, 165, and 153 passengers for Air Canada, Delta Airlines, ITA Airways, and Turkish Airlines, respectively. We therefore assume that our composite cabin for the Airbus A320-200 has a seating capacity of 155.

For each cabin class *j* in aircraft *i*, we determine the number of seats in each scenario using the following procedure:

$$No. of\ Seats_{ij} = \frac{Proportion\ of\ Floor\ Space_{ij}\ \times Total\ Floor\ Space\ (Sq.Ft.)_i}{Sq.Ft.Per\ Seat_{ij}}.$$

Emissions Estimation: Emissions estimates are derived by calculating the fuel requirements for the four aircraft types leveraged by our model (i.e., Airbus A320-200, Boeing 737-800, Airbus A330-200, and Boeing 777-200LR). When doing so, we used the previously derived composite cabin to inform passenger capacity in each class of travel for each aircraft type. For example, our composite cabin for the Airbus A320-200 has a passenger seating capacity of 155, with 131, 14, and 11 seats available in economy, premium economy, and business, respectively.[2] We use these figures when deriving the Airbus A320-200's fuel requirements. In addition to assuming complete occupancy across all cabins (and estimating fuel requirements on that basis), fuel estimates are also derived for various load factors (percentage of seats that are full), as doing so enables us to ascertain the extent to which the number of passengers on board impacts fuel requirements (and, consequently, emissions). For each aircraft type, we calculate fuel requirements for both short-haul and long-haul flights. A total of 47 routes (24 short-haul and 23 long-haul) inform fuel estimates, with an average being computed for each stage length (20, 21). Emission estimates are derived by converting fuel requirements into carbon emissions, assuming the production of 3.15 kg $CO_2$ per kilogram of kerosene burnt (22).

Aircraft emissions are estimated by accounting for departure and arrival points, the number of passengers on board, and the corresponding fuel consumption per flight.[3] When modeling emissions, long-haul routes could introduce the possibility of an increasing marginal fuel burn per passenger. Consequently, these routes generate a higher emissions penalty per passenger, as additional passengers require more fuel, and transporting that fuel imposes a further penalty. However, both the existing literature and our acquired dataset indicate that, for both short and long-haul flights, the marginal emissions penalty per passenger remains constant, implying a linear relationship between the added weight on an aircraft and the corresponding fuel required to carry that weight (23). We perform linear regression analyses on the dataset for both short and long-haul flights for each of the four aircraft (8 regressions in total), using the following equation:

$$Emissions\ Per\ Flight_i = \beta_o + PAX\ Emissions \times No.of\ Passengers_i + \epsilon_i,$$

where $Emissions\ Per\ Flight_i$ = total emissions produced from flight $i$, $\beta_o$ = emissions associated with flying the aircraft without passengers (i.e., emissions for flying an empty aircraft shell and seats within the baseline seating configuration), $PAX\ Emissions$ = emissions generated from the fuel required to carry an additional passenger and their luggage, $No.of\ Passengers_i$ = number of passengers on flight $i$, and $\epsilon_i$ = error term (for results, see Supplementary Information: Table S2e).

Next, we calculate the standardized emission constant associated with the added weight on the aircraft. We use the following equation:

$$Emissions\ factor\ \left(\frac{kg\ CO_2}{kg\ Weight}\right) = \frac{PAX\ Emissions}{PAX\ Weight},$$

where $PAX\ Emissions$ = emissions generated from the fuel required to carry an additional passenger and their luggage, and $PAX\ Weight\ (kg)$ = weight of a passenger and their luggage (we assume a passenger weighs 65 kg and their luggage weighs 10 kg). Although the emissions constant is calculated using emissions generated from the fuel required to carry an additional passenger and their luggage, it can be applied to seat-related emissions as well, since the constant reflects the emissions from additional weight on the aircraft – regardless of whether that weight comes from seats, passengers, or luggage. To calculate the emissions associated with carrying the seats (assuming that a business class seat weighs 140 kg, a premium economy seat weighs 20 kg, and an economy seat weighs 10 kg), we use the following formula:

$$Seating\ Emissions\ (kg\ CO_2)_{ij} = No.of\ Seats_{ij} \times Weight\ Per\ Seat\ (kg)_j \times Emissions\ Factor\ (kg\ CO_2/kg\ Weight)_i.$$

---

[2] Cabin-level seating capacity may not sum to the total aircraft seating capacity due to rounding errors. However, our model employed the un-rounded values in all calculations.

[3] We estimate this relationship by enumerating fuel consumption at different load factors for the aircraft and routes in our database. We subsequently calculate fuel consumption with no passengers, and the marginal increase in fuel consumption for each additional passenger (see Supplementary Information: Table S2a-S2d). Fuel consumption is estimated at 0, 50, and 100 percent load factors. In cases where 100 percent load factor exceeds the maximum allowed by the emissions calculator utilized, we use the maximum allowed passenger number of 300.

In our $Emissions\ Per\ Flight_i$ equation, $\beta_o$ captures the emissions for flying without passengers. To isolate the emissions attributed solely to the aircraft body (Empty Aircraft Emissions), we use the following equation:

$$Empty\ Aircraft\ Emissions_i = \beta_o - \sum_j Seating\ Emissions_{ij},$$

where $Empty\ Aircraft\ Emissions_i$ are the fixed emissions generated from the fuel burn required to carry an empty aircraft $i$, and $Seating\ Emissions_{ij}$ are the emissions generated from carrying the seats in the cabin $j$. For each aircraft, we are thus able to estimate the fixed emissions generated from the fuel required to carry an empty aircraft, and the emissions constant to describe the emissions generated from the fuel needed to carry the seats and the additional weight of each passenger and luggage, based on the inputted class distribution (Table 1).

Next, we calculate the emissions associated with carrying an additional passenger and their luggage using the following equation:

$$PAX\ Emissions\ (kg\ CO_2)_{ij} = No.of\ Seats_{ij} \times PAX\ Weight\ (kg) \times Emissions\ Factor\ (kg\ CO_2/kg\ Weight)_i.$$

For each aircraft and scenario, we estimate aggregate emissions per flight using the following equation (all units in kg $CO_2$):

$$Emissions\ Per\ Flight_i = Empty\ Aircraft\ Emissions_i + \sum_j Seating\ Emissions_{ij} + \sum_j PAX\ Emissions_{ij}.$$

To calculate the total emissions per passenger in cabin $j$, we account for two components: 1) the emissions associated with the passenger's cabin square footage, and 2) an evenly distributed share of the service zones. We use the following equation:

$$Emissions\ Per\ Passenger_{ij} = \frac{Emissions\ Per\ Flight_i \times Proportion\ of\ Floor\ Space_{ij}}{No.of\ Seats_{ij}} + \frac{Emissions\ Per\ Flight_i \times Proportion\ of\ Floor\ Space\ for\ Service\ Zones}{No.of\ Seats_i}.$$

Finally, using the maximum lifetime number of flight cycles and flight hours for each aircraft, we determine the maximum number of flights an aircraft can make. We then calculate the lifetime emissions of each aircraft in each scenario using the following formula:

$$Lifetime\ Emissions_i = Emissions\ Per\ Flight_i \times Max.No.of\ Flights_i.$$

We calculate figures for both total emissions and emissions solely associated with transporting passengers and luggage (i.e., excluding the fixed emissions associated with carrying the aircraft itself). We generate figures for narrow-body aircraft by averaging the computed values for the A320-200 and 737-800. We use a similar approach for wide-body aircraft, which use data for the A330-200 and 777-200LR.

Revenue Generation: To estimate revenue contributions for each cabin, we source publicly available airfare data for short-haul and long-haul routes flown by 12 global airlines. These airlines typically offer three different classes of travel (i.e., business, premium economy, and economy) on the same routes used in our emissions estimates. Airfare data for these routes is sourced directly from the airline's website, and the selected routes have average travel distances of 631 nautical miles and 3,003 nautical miles established prior. We recognize that airfares are constantly changing, reflecting dynamic shifts in supply and demand, seat availability, and evolving airline revenue management strategies. To account for these changes, airfare data is sourced for three months: August 2025, November 2025, and February 2026. This data collectively yields 90 days of airfare information and ensures, we argue, that airfare data in our model is less susceptible to one-off events that may impact airfares (e.g., adverse weather conditions).

The collection of airfare data for two short-haul and two long-haul routes, each served by 12 airlines offering three classes of travel, yields 144 airfares. This figure rises to 12,816 airfares after accounting for three months' worth of data. We focus on routes operated with a 3-class cabin configuration (see Supplementary Information: Table S3a and S3b, for 90-day rolling averages).

Airfare data is subsequently aggregated across airlines and periods to yield separate airfare estimates for short-haul and long-haul travel stratified by travel class (Table 1). In sourcing airfare data, we note that our approach does not allow us to control for aircraft type for every airline's complete roster of short-haul and long-haul flights. For example, Emirates and British Airways both fly the London-Dubai route but do so using different aircraft types. However, we do not believe this discrepancy significantly alters our revenue estimates, as airfares generally do not vary based on aircraft type, but rather on airline, stage length, and travel class. Our analysis accounts for these parameters.

For each scenario, assuming static ticket airfares, we calculate the revenue from each flight as follows:

$$Revenue\ Per\ Flight\ (\$)_i = \sum_j Number\ of\ Seats_{ij} \times Ticket\ Price\ (\$)_i.$$

We then determine the required economy airfare needed to maintain constant total revenue after removing premium travel cabins (i.e., business and premium economy) using the following formula:

$$\text{New Ticket Price (\$)}_{ij} = \frac{\text{Baseline Revenue Per Flight}_i}{\text{Revised No. of Seats}_{ij}},$$

where $\text{Baseline Revenue Per Flight}_i$ = total revenue generated per flight under the original cabin configuration, $\text{Revised No. of Seats}_{ij}$ = total number of economy seats after replacement of premium cabins with economy, and $\text{New Ticket Price (\$)}_{ij}$ = economy airfare required under the revised configuration to preserve baseline revenue.

We then calculate the lifetime revenue of each aircraft in each scenario using the following formula:

$$\text{Lifetime Revenue}_i = \text{Revenue Per Flight}_i \times \text{Max. No. of Flights}_i.$$

Scenarios Considered: In our analysis, we consider two scenarios: a baseline scenario with a default 3-cabin layout and a scenario in which we assume an all-economy-class configuration. We select an all-economy configuration for comparison as it allows us to focus on the impact of eliminating premium travel (as opposed to a scenario where some premium seating is maintained, albeit with a smaller footprint). The aircraft seating configurations leveraged by our model reflect this choice (Table 2). We further enumerate the number of seats, as well as the weight of seats, passengers, and luggage for each composite narrow-body and wide-body aircraft across short-haul and long-haul routes (Table 3). We subsequently enumerate the emissions factors (kg $CO_2$/kg weight) for additional weight beyond the empty aircraft frame (Table 4). We note a slight difference in emissions factors between scenarios due to aggregation, as the results for the composite aircraft are calculated by averaging the results of each model (such as the A320-200 and 737-800 for the narrow-body aircraft). However, since moving from a 3-cabin to an all-economy configuration changes the total number of seats on each aircraft – and therefore the total onboard weight of passengers and seats – each aircraft's emissions factor is multiplied by a weight that reflects its seating configuration when calculating total emissions.

**Results and Discussion**

What is the impact of foregoing premium air travel? We address this question by examining how refurbishing a 3-class cabin (economy, premium economy, and business) to an all-economy configuration affects emissions associated with flying. In scrutinizing emissions, we consider – for both short-haul and long-haul routes – emissions per passenger, emissions per flight, and emissions generated over the aircraft's lifespan. We calculate emissions in two ways: first, we consider the total emissions required to transport the empty aircraft, as well as seats, passengers, and luggage. Second, noting that a significant proportion of an aircraft's weight is the fixed weight of the empty airframe, we also repeat this analysis to consider only 'variable' emissions, defined as the emissions solely associated with transporting seats, passengers, and luggage (i.e., absent the aircraft itself). In addition to scrutinizing the emissions impact of foregoing premium travel, we also consider how this may impact revenue. Here, we examine the effects of eliminating a 3-class cabin on aggregate revenue per flight and revenue generated across the aircraft's lifespan. Furthermore, we assess how changes in revenue generated on a per-flight basis may impact average ticket prices.

Our findings are threefold.

First, we find that on a per-passenger basis, there is a significant decline in emissions when comparing an all-economy configuration to a 3-class configuration. Here, we compare emissions attributed to an economy-class passenger in a 3-class configuration to those in an all-economy configuration. Specifically, we observe that foregoing premium travel classes on narrow-body aircraft reduces per-economy class passenger emissions by 8.36 percent and 8.05 percent on short-haul and long-haul flights, respectively. On wide-body aircraft, we observe more significant declines, as emissions per economy-class passenger decrease by 21.52 percent and 20.86 percent on short-haul and long-haul flights, respectively (Fig. 2a). As expected, an all-economy configuration for a much larger number of passengers leads to a substantial reduction in emissions per passenger. More sizeable reductions occur for wide-body aircraft due to the larger increase in capacity for economy-class passengers.

Our second finding is that both on a per-flight basis and over the aircraft's lifespan, there is minimal change in emissions when moving from a 3-class cabin to an all-economy configuration (Fig. 2b-2c). On a per-flight basis, for narrow-body aircraft, emissions on short-haul and long-haul flights decrease by an average of 0.31 percent and 0.39 percent, respectively, in an all-economy configuration (compared to a 3-class cabin). For wide-body aircraft, we observe a slight increase in emissions on a per-flight basis, with gains of 1.12 percent and 1.43 percent for short-haul and long-haul flights, respectively. We observe similar results when aggregating the lifetime emissions of the various aircraft types. For narrow-body aircraft, lifetime emissions decrease by 0.36 percent and 0.45 percent on short-haul and long-haul flights, respectively. On wide-body aircraft, lifetime emissions increase by 1.11 percent and 1.43 percent on short-haul and long-haul flights, respectively. Importantly, per-flight and aggregate emissions, rather than per-passenger emissions, provide – we argue - a more complete measure of aviation's climate impact.

What explains directional heterogeneity between narrow-body versus wide-body aircraft emissions? We note that on a wide-body aircraft, an all-economy seating configuration allows for a proportionally higher number of passengers than a narrow-body aircraft. The increase in emissions from those additional passengers offsets the decline in emissions from eliminating premium cabins and the heavier seats associated with them. For example, we find that on a wide-body aircraft in a 3-cabin configuration, the seats alone weigh 7,042 kg, while passengers and luggage weigh 19,697 kg (Table 3). While moving to an all-economy configuration results in the total weight of seats declining by 3,592 kg due to the elimination of premium class seats (corresponding to emissions declining by 573 kg on short-haul and 2,242 kg on long-haul flights), the increase of the weight of passengers and luggage by 6,181 kg (corresponding to emissions increasing by 986 kg on short-haul aircraft and 3,858 kg on long-haul aircraft) disproportionally offsets this. We note that the slight change in emissions factor, as previously explained, contributes to a decrease of 5 kg in emissions on short-haul flights and 16 kg on long-haul flights, resulting in a total net increase of 408 kg and 1,600 kg in emissions on wide-body short-haul and long-haul flights, respectively.

We note that similar results are observed – albeit more profoundly – when the aircraft's empty weight is excluded. Considering only 'variable emissions,' that is, emissions associated solely with transporting passengers, crew, and luggage, we observe that for narrow-body aircraft, variable emissions per flight decrease by 1.70 percent and 1.79 percent on short-haul and long-haul flights, respectively. On wide-body aircraft, variable emissions per flight increase by 9.58 percent and 9.59 percent on short-haul and long-haul flights, respectively (Fig. 3a). Similar results hold when calculating lifetime emissions – on narrow-body aircraft, lifetime variable emissions decrease by 1.97 percent and 2.06 percent on short-haul and long-haul flights, respectively. On wide-body aircraft, lifetime variable emissions increase by 9.62 percent and 9.61 percent on short-haul and long-haul flights, respectively (Fig. 3b). Thus, when considering just the variable component, we observe more significant changes in emissions[4].

Third, we observe significant impacts on ticket prices, revenue per flight, and revenue over an aircraft's lifespan. To illustrate the financial implications of switching to an all-economy cabin, we calculate the amount that economy class ticket prices would need to adjust by to achieve the same revenue as the baseline scenario (Fig. 4a). This figure may be interpreted as the amount that economy class tickets may rise by if premium cabins were eliminated, or the subsidy amount that the government may have to provide to incentivize such a configuration. For narrow-body aircraft, the average economy-class ticket price must increase by $13.24 (5.99%) for short-haul flights and rise by $153.14 (29.98%) for long-haul flights. For wide-body aircraft, ticket prices must fall by $10.37 (-4.69%) and increase by $124.31 (24.34%) for short-haul and long-haul flights, respectively. All percentage changes are relative to prices in the baseline configuration.

What impact would such an increase have on the average consumer? According to a survey of American consumers, the average American sets a per-trip budget of $2,743 (28). While an increase in ticket prices of $13.24 on narrow-body aircraft flying short-haul flights might represent just 0.481 percent of this budget, an increase by $124.31 on wide-body aircraft flying long-haul flights represents 4.52 percent of this budget – a noteworthy change.

---

[4] Here, we do not consider variable emissions per passenger. Since emissions are calculated based on weight, variable emissions are the same for an all-economy passenger in a 3-cabin configuration and an all-economy configuration.

Considering differing demand elasticities may also provide additional insight into the impact of raising prices on air travel consumption. A meta-analysis of the average price elasticities of international tourism demand found that tourists from Asia are the most price-sensitive, with an average price elasticity of demand of -1.420, followed by tourists from America with an average price elasticity of demand of -1.277. A 5.99 percent increase in the price of narrow-body short-haul flights would lead to a 7.09 percent decrease in consumption of such flights for Asian tourists and a 5.37 percent decrease for American tourists, for instance. Meanwhile, African tourists have the smallest price elasticity of -0.783, suggesting that they are less price-sensitive than consumers from other regions (29).

We note that the increased airfares that result from eliminating premium class options – especially on wide-body, long-haul flights – risk imposing a regressive burden on working-class families. While eliminating premium classes may, in principle, democratize aircraft seating, the ensuing increase of up to 24.3 percent in airfare could have the opposite effect. With markedly higher price elasticities of demand compared to wealthier passengers, the required airfare increases to maintain revenue parity are more likely to exclude working-class consumers than more affluent households. Indeed, as airfare increases have outpaced inflation in the past year, 68 percent of leisure travelers have concurrently reported heightened price sensitivity (30). Considering travel a luxury good, working-class households are unlikely to allocate the rising shares of their earnings to travel that airfare increases would necessitate, opting instead to forego air travel entirely. While the subsequent reduction in flight demand may, in the long term, decrease the number of flights approved by airlines and reduce aviation emissions, in the near term, we anticipate that flights will continue to operate despite low passenger load factors (as empirically observed). The starkest consequence of eliminating premium options is therefore the inadvertent removal of affordable travel options.

Accordingly, due to subsequent airfare increases, from a socioeconomic perspective, we find that the removal of premium classes may inadvertently lead to a disproportionate weakening of working-class families' access to air travel. Currently, airlines offering premium travel classes operate as a form of second-degree price discrimination, providing different "versions" of air travel to consumers according to their willingness to pay. Employing this strategy allows airlines to redirect additional revenue derived from affluent travelers to cross-subsidize lower airfares for economy passengers. In doing so, airlines increase allocative efficiency and total welfare; removing premium cabins, and by extension this system, would drastically undermine these redistributive mechanisms. Effectively, by paying for premium classes, wealthy travelers ensure that their less affluent counterparts have access to air travel.

Therefore, a policy shift away from the 3-cabin layout risks threatening air travel equity. Studies of current aviation taxation reveal that imposing broad-based travel costs functions as a regressive policy. A survey of the impact of the UK's Air Passenger Duty, a tax levied on passengers departing from UK airports, reported that even seemingly modest additional charges, such as a £52 levy for a family of four, can amount to up to 28 percent of a working family's weekly income – serving as a substantial deterrent from air travel (31). In a similar vein, removing premium cabins and imposing cost burdens on economy flyers through increased economy airfares would likely have a similar effect: diminishing access to air travel for economically disadvantaged groups. This outcome threatens to undermine aviation's current gains in equity and social welfare, which have been achieved using premium classes by wealthy travelers and have ensured greater accessibility of air travel for working families.

Additionally, eliminating premium class cabins risks alienating a disproportionately high yield subsegment of airlines' passenger markets: wealthier travelers. These travelers, whose willingness to pay exceeds that of the average consumer and who exhibit markedly more inelastic price elasticities of demand, despite constituting only 12 percent of the total passenger volume, generate over 50 percent of total passenger revenue (13). U.S. households with earnings of $100,000 or more account for 75 percent of air travel spending and emit 13.74 t$CO_2$ per year in transport emissions (32, 33). Long-haul routes magnify this yield asymmetry, with our analysis finding that premium economy airfares and business class airfares command 2.28 times and 5.40 times the price of economy airfares, respectively. This price disparity is primarily due to the premium classes' ability to maximize consumer surplus across heterogeneous demand profiles. Thus, due to the substitution effect, the lost revenue associated with the removal of premium classes may even be understated in our model.

Furthermore, the absence of premium classes risks potentially inducing substitution effects amongst high-income travelers. Premium travel currently serves as an intermediate offering between standard economy and alternative modes of transportation, targeting wealthy passengers who prefer both the perks associated with premium travel and its comparably lower cost as an alternative to other methods of transport (e.g., private aviation) (33). Without this intermediate offering, many wealthier passengers, due to their preference for high-end travel outweighing the increased cost of non-commercial carriers, may bypass airline travel entirely. In an industry with structurally thin margins, such defections are economically significant, especially when accounting for the stabilizing effect of these passengers. Indeed, during periods of economic turbulence, business-oriented premium passengers provide an earnings buffer, helping to mitigate or offset decreases in broader overall demand. For instance, in the June 2025 quarter, United Airlines' premium revenue rose 5.6 percent compared to the same quarter a year ago, while overall passenger revenue increased by just 1.1 percent (33). By removing premium classes, airlines risk eroding high-yield demand and dismantling a key insulator against downturn-induced crises.

The defection of wealthy passengers to private aviation has the additional consequence of worsening emissions, on both a per-flight and per-passenger basis. Private aircraft, which typically have a significantly lower load factor (often fewer than 10 passengers per flight), result in a fixed fuel burden distributed over fewer passengers, thereby increasing the per-passenger emissions of these flights. These aircraft operate only when commissioned as opposed to their scheduled commercial counterparts, further inflating per-passenger carbon intensity (34). Estimates of average fuel use per passenger hour for private aviation vary but are typically between 10 and 20 times higher than the average fuel use per passenger hour for a commercial flight. The implications for the aviation sector's emissions are severe: the early pandemic's 20 percent increase in the number of private aviation flights in the U.S. resulted in a 23 percent increase in $CO_2$-equivalent emissions (35). Therefore, policies that remove premium classes to mitigate aircraft emissions may paradoxically, due to private aircraft substitution, have the opposite effect.

We also observe significant impacts on revenue per flight and across the aircraft's lifespan, though these results differ based on aircraft type and route. Holding ticket prices constant, we observe declines in revenue across the majority of aircraft types and route lengths when switching to an all-economy configuration (Fig. 4b-4c). On narrow-body aircraft, revenue per flight falls by 5.66 percent and 23.08 percent on short-haul and long-haul flights, respectively. On wide-body aircraft, revenue per flight increases by 4.92 percent on short-haul flights but decreases by 19.57 percent on long-haul flights (Fig. 4b). Similar results are seen across the aircraft's lifespan. On narrow-body aircraft, revenue per flight falls by 5.84 percent and 23.44 percent on short-haul and long-haul flights, respectively. On wide-body aircraft, revenue per flight increases by 5.03 percent on short-haul flights but falls by 19.42 percent for long-haul flights (Fig. 4c).

This disparity in results – increasing or decreasing of per-flight revenue – can be attributed to the differences in ticket prices across cabin classes on both short-haul and long-haul routes. On short-haul routes, ticket prices in business and premium economy are, on average, 2.72 and 1.56 times higher than those in economy, respectively. On long-haul routes, these multiples rise to 5.40 and 2.28, respectively. Hence, on wide-body aircraft, the revenue from a 3-class cabin configuration on short-haul routes is equivalent to that of 329 economy-class seats, below the seating capacity of 345 seats in the all-economy scenario. On long-haul routes, however, the revenue from a 3-class cabin configuration is equivalent to that of 431 seats, greater than the seating capacity in an all-economy scenario. Thus, in the long-haul scenario, an all-economy configuration for wide-body aircraft generates lower revenue.

In sum, transitioning commercial aircraft from a 3-class to an all-economy configuration would result in a significant change in emissions on a per-passenger basis, but – due to the high fixed level of fuel required to transport an empty aircraft – produces only marginal changes in emissions per flight and over the aircraft's lifespan. The revenue effects are more pronounced, particularly on long-haul routes, where the loss of high-yield premium seating substantially reduces total revenue. We note, however, that short-haul flights are the most likely candidates for conversion to an all-economy configuration, as revenue either decreases marginally (in the case of narrow-body aircraft) or remains unchanged (in the case of wide-body aircraft).

**Limitations and Conclusion**

To assess the impact of eliminating premium travel, we estimate the marginal emissions change on a per-passenger, per-flight, and operational lifespan basis. Concurrently, we quantify the change in revenue associated with an all-economy configuration and the compensatory adjustments in airfare required to sustain baseline revenue under this configuration. Our analysis reveals that, on a per-passenger basis, emissions uniformly decline, reflecting the benefits of improved seat density. However, this decline is not uniformly observed on a per-flight and operational lifespan basis. Here, we observe heterogeneity in the impact of emissions according to aircraft class, with wide-body aircraft experiencing a net increase in lifetime total emissions, while narrow-body aircraft exhibit modest reductions. Furthermore, we find that across most aircraft and route types, revenue losses are associated with eliminating a 3-cabin structure, necessitating substantial increases in ticket prices. These increases, in turn, may constrain access to air travel for price-sensitive demand segments and attenuate overall demand for commercial air travel.

To our knowledge, this study is the first to quantify the emissions and financial impact of eliminating premium travel classes. While few would dispute that premium travel is more polluting on an individual basis, we highlight the importance of understanding the overall impact on emissions and revenue to determine the feasibility of such a measure. We acknowledge, however, that our study has several limitations.

First, we assume a 100 percent load factor across cabins in our study to isolate the impact of eliminating premium classes of travel. However, we recognize that load factors may fall short of 100 percent and may vary between cabins (36, 37). Additionally, the compounding effect of travel class and load factor may exaggerate a passenger's carbon footprint (e.g., low load factors in premium cabins can exacerbate the emissions footprint of premium passengers, as aggregate cabin emissions are divided among fewer passengers, resulting in higher per-passenger emissions) (10). We note, however, that this should not change the directionality of our aggregate results. Moreover, driven by higher revenue potential and corporate traveler preferences, airlines are increasingly leveraging pathways to sell more premium seats, bringing the observed load factors for these cabins into parity with those seen in economy class.

Second, we assume airlines currently possess a 3-cabin structure, a configuration not observed by budget airlines (airlines that exclusively offer economy seating). This decision is based in large part on the fact that budget airlines, commonly referred to as low-cost carriers (LCCs), hold a minority share of the aviation market. Of the 811 airlines worldwide, only 102 are low-cost carriers, representing less than 13 percent. Additionally, LCCs operate only 30 percent of all scheduled flights and 33 percent of all scheduled airline seats, rendering them a substantial yet distinct minority in the aviation industry (38). Moreover, it is important to note that these LCCs are currently transitioning away from their traditional all-economy model toward quasi-hybrid configurations that incorporate premium offerings, aligning them with the conventional airlines we analyze. Notably, airlines such as Spirit, Southwest, and Frontier have begun to integrate premier options such as extra-legroom seating and bundled airfare tiers as a response to declining revenue; Spirit Airlines has lost more than $2.2 billion since the start of 2020, while Frontier has not reported a full-year profit since 2019 (39). The retreat away from their all-

economy model by LCCs, therefore, may serve as further evidence that premium classes are not merely luxuries, but structural necessities in contemporary airline economics.

Third, while revenue impacts constitute a central component of our analysis, we acknowledge that they may not be the only determinant of airline decision-making. Indeed, transitioning from a premium class structure to an all-economy configuration may result in lower production costs per aircraft. Nevertheless, we note that revenue generated from seat sales alone is generally insufficient to ensure profitability, with empirical evidence reporting that all U.S. airlines would be operating at a loss absent loyalty programs (40). Moreover, given that premium classes disproportionately contribute to airline yield, their removal may further compress already slim operating margins.

Fourth, we acknowledge a methodological limitation in our approach to revenue estimation. By collecting airfare data from a single route segment, our analysis may not accurately capture the total revenue yield of a flight, as airfare is often determined on a network-wide basis. This whole-of-network pricing framework considers connecting flights and incorporates factors that a segment-level analysis cannot fully account for. However, given the complexity associated with network pricing, we have chosen a segment-based methodology as the best estimate of flight-level revenue.

Finally, our analysis proceeds under the assumption that aggregate demand is sufficient to support a hypothetical all-economy configuration. In practice, however, we acknowledge that economy class seats may be an imperfect substitute for the demand currently met by premium class seating. Therefore, eliminating these premium cabins, particularly first-class offerings, risks displacing high-yield passengers towards alternative travel methods, notably private aviation (41). Such a substitution risks both diminishing commercial airlines' revenue and exacerbating per-passenger emissions.

Limitations notwithstanding, our study enumerates – to our knowledge, for the first time – novel emissions and economic consequences of foregoing premium travel classes. We observe that all-economy travel, compared to premium inclusive configurations, yields marginal emissions benefits – and in some instances, either no benefit at all or a net increase in emissions. More importantly, we find profound economic consequences of all-economy travel. Here, the burden of increased airfare that a shift to an all-economy shift levies falls most heavily on working-class travelers, as removing premium cabins erodes the current cross-subsidization that airlines currently rely upon to ensure the accessibility of travel for working-class travelers. Therefore, paradoxically, by paying for premium travel, affluent travelers underwrite systemwide air travel access. The removal of these classes thus poses substantial threats to the equity and inclusivity of commercial air travel.

| | Aircraft | | A320-200 | | 737-800 | | A330-200 | | 777-200LR | |
|---|---|---|---|---|---|---|---|---|---|---|
| | Stage Length | | Short Haul | Long Haul | Short Haul | Long Haul | Short Haul | Long Haul | Short Haul | Long Haul |
| Emissions | Empty Aircraft Emissions (kg $CO_2$) | | 12,622.64 | 39,283.16 | 13,412.04 | 41,873.84 | 29,101.76 | 86,474.52 | 35,007.68 | 103,645.76 |
| | Emissions Factor (kg $CO_2$/ kg Weight) | | 0.20 | 0.73 | 0.19 | 0.76 | 0.18 | 0.68 | 0.14 | 0.58 |
| Cabin Space Allocation | Cabin Seating Quantities (#) | Economy | 131.34 | 131.34 | 133.95 | 133.95 | 179.33 | 179.33 | 245.49 | 245.49 |
| | | Premium Economy | 13.37 | 13.37 | 9.00 | 9.00 | 17.46 | 17.46 | 18.35 | 18.35 |
| | | Business | 10.54 | 10.54 | 14.62 | 14.62 | 30.3 | 30.3 | 34.85 | 34.85 |
| | Cabin Allocation (%) | Economy | 51 | 51 | 49 | 49 | 35 | 35 | 38 | 38 |
| | | Premium Economy | 6 | 6 | 4 | 4 | 5 | 5 | 4 | 4 |
| | | Business | 9 | 9 | 12 | 12 | 19 | 19 | 18 | 18 |
| | | Service Zones | 33 | 33 | 36 | 36 | 42 | 42 | 40 | 40 |
| | Cabin Spatial Allocation (sq ft) | Economy | 2,468.08 | 2,468.08 | 2,616.79 | 2,616.79 | 3,812.99 | 3,812.99 | 4,851.69 | 4,851.69 |
| | | Premium Economy | 312.06 | 312.06 | 190.01 | 190.01 | 555.49 | 555.49 | 539.79 | 539.79 |
| | | Business | 456.29 | 456.29 | 620.51 | 620.51 | 2,036.97 | 2,036.97 | 2,293.29 | 2,293.29 |
| | | Service Zones | 1,579.33 | 1,579.33 | 1,887.59 | 1,887.59 | 4,605.20 | 4,605.20 | 5,091.24 | 5,091.24 |
| | Seating Spatial Allocation (sq ft) | Economy | 18.79 | 18.79 | 19.54 | 19.54 | 21.26 | 21.26 | 19.76 | 19.76 |
| | | Premium Economy | 23.34 | 23.34 | 21.11 | 21.11 | 31.82 | 31.82 | 29.41 | 29.41 |
| | | Business | 43.31 | 43.31 | 42.44 | 42.44 | 67.23 | 67.23 | 65.81 | 65.81 |
| | | Service Zones | 10.215 | 10.215 | 12.01 | 12.01 | 20.16 | 20.16 | 17.345 | 17.345 |
| Financial | Airfare Price ($) | Economy | 221.02 | 510.8 | 221.02 | 510.8 | 221.02 | 510.8 | 221.02 | 510.8 |
| | | Premium Economy | 343.92 | 1,162.53 | 343.92 | 1,162.53 | 343.92 | 1,162.53 | 343.92 | 1,162.53 |
| | | Business | 601.10 | 2,757.49 | 601.10 | 2,757.49 | 601.10 | 2,757.49 | 601.10 | 2,757.49 |

|  | Flight Revenue ($) | 39,959.97 | 111,682.49 | 41,488.31 | 119,197.82 | 63,853.06 | 195,450.46 | 81,515.34 | 242,818.17 |
|---|---|---|---|---|---|---|---|---|---|
| Other | Flight Distance (nm) | 631.00 | 3002.67 | 631.00 | 3002.67 | 631.00 | 3002.67 | 631.00 | 3002.67 |
|  | Max Flight Cycles | 48,000 | 48,000 | 55,000 | 55,000 | 40,000 | 40,000 | 60,000 | 60,000 |
|  | Max Flight Hours | 60,000 | 60,000 | 90,000 | 90,000 | 130,000 | 130,000 | 160,000 | 160,000 |

*Table 1: Input Parameters by Aircraft and Route Type*

|  | Aircraft | Cabin | Baseline Scenario | All-Economy Scenario |
|---|---|---|---|---|
| Narrow body | A320-200 | Economy | 131.34 | 172.23 |
|  |  | Premium Economy | 13.37 | 0.00 |
|  |  | Business | 10.54 | 0.00 |
|  | 737-800 | Economy | 133.95 | 175.44 |
|  |  | Premium Economy | 9.00 | 0.00 |
|  |  | Business | 14.62 | 0.00 |
| Wide body | A330-200 | Economy | 179.33 | 301.25 |
|  |  | Premium Economy | 17.46 | 0.00 |
|  |  | Business | 30.30 | 0.00 |
|  | 777-200LR | Economy | 245.49 | 388.84 |
|  |  | Premium Economy | 18.35 | 0.00 |
|  |  | Business | 34.85 | 0.00 |

*Table 2: Aircraft Seating Configurations by Scenario*[5]

---

[5] All-Economy Scenario seating estimates are lower than the certified exit limit for each respective aircraft (24-27).

|  |  | Baseline 3-Cabin Scenario | | | All-Economy Scenario | | |
|---|---|---|---|---|---|---|---|
|  |  | Seat Quantity (#) | Seat Weight (kg) | Passenger and Luggage Weight (kg) | Seat Quantity (#) | Seat Weight (kg) | Passenger and Luggage Weight (kg) |
| Narrow-body | Economy | 132.64 | 1,326.40 | 9,948.00 | 173.83 | 1,738.30 | 13,037.25 |
| | Premium Economy | 11.19 | 223.80 | 839.25 | 0.00 | 0.00 | 0.00 |
| | Business | 12.58 | 1,761.20 | 943.50 | 0.00 | 0.00 | 0.00 |
| | Total | 156.41 | 3,311.40 | 11,730.75 | 173.83 | 1,738.30 | 13,037.25 |
| Wide-body | Economy | 212.41 | 2,124.10 | 15,930.75 | 345.04 | 3,450.40 | 25,878.00 |
| | Premium Economy | 17.91 | 358.20 | 1,343.25 | 0.00 | 0.00 | 0.00 |
| | Business | 32.57 | 4,559.80 | 2,422.75 | 0.00 | 0.00 | 0.00 |
| | Total | 262.89 | 7,042.10 | 19696.75 | 345.04 | 3,450.40 | 25,878.00 |

*Table 3: Weight Inputs for Different Aircraft Types*

|  |  | Baseline 3-Cabin Scenario | All-Economy Scenario |
|---|---|---|---|
| Narrow-body | Short Haul | 0.15954 | 0.15938 |
| Narrow-body | Long Haul | 0.62422 | 0.62366 |
| Wide-body | Short Haul | 0.19445 | 0.19458 |
| Wide-body | Long Haul | 0.74762 | 0.74747 |

*Table 4: Emissions Factors (kg $CO_2$/kg Seats, PAX and Luggage Weight) for Different Scenarios*

*Note: Numbers are listed to 5 significant figures as the marginal difference in emissions factors accounts for a small proportion of the change in emissions moving from the baseline 3-cabin to the all-economy scenario.*

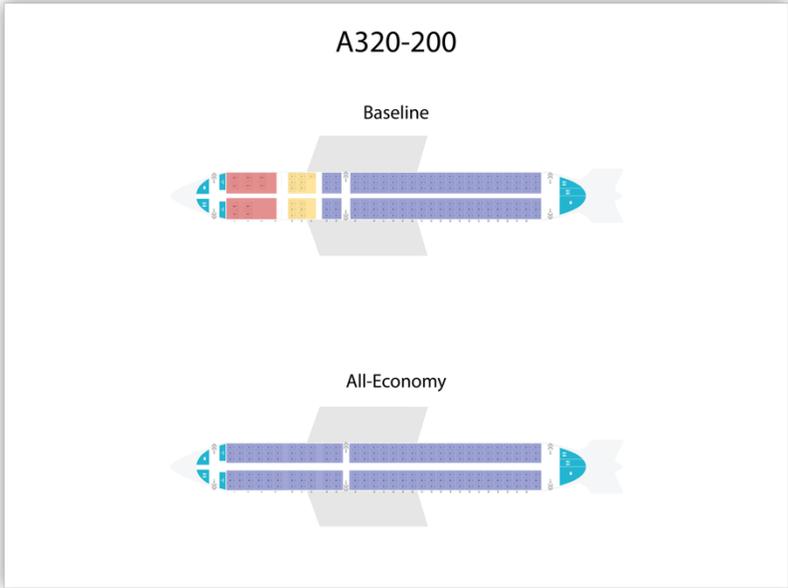
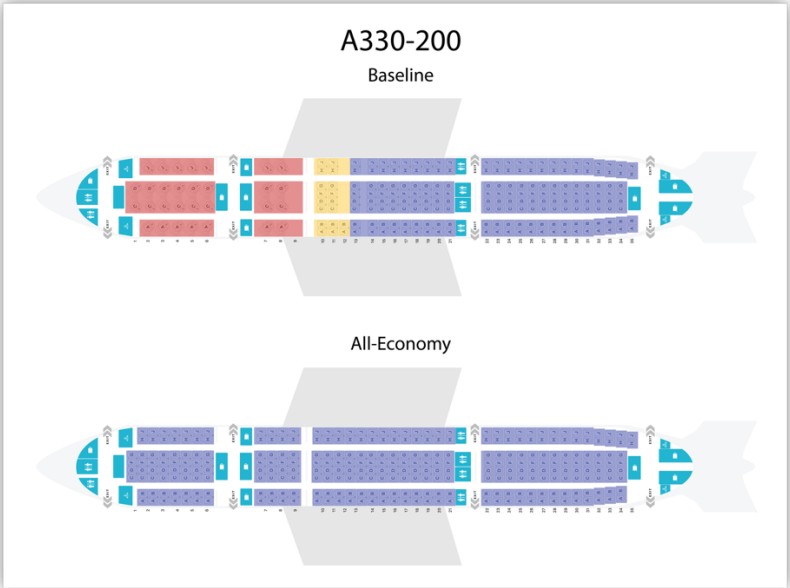
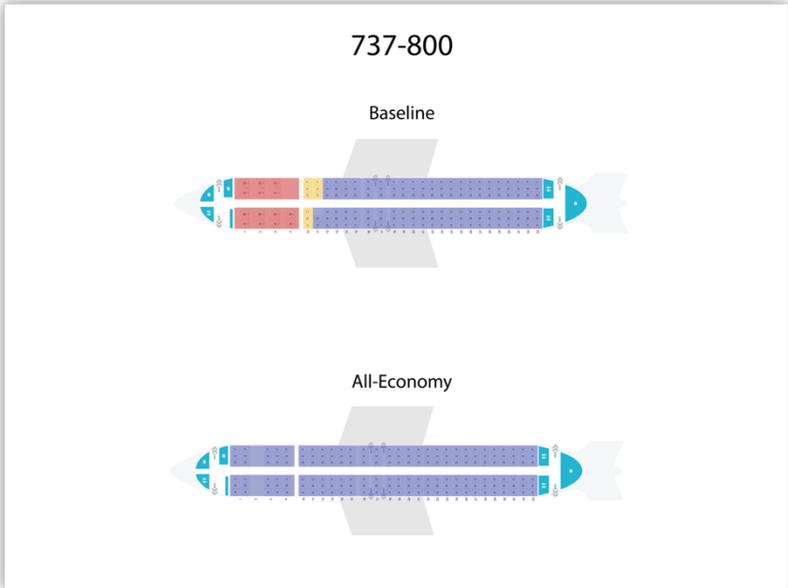
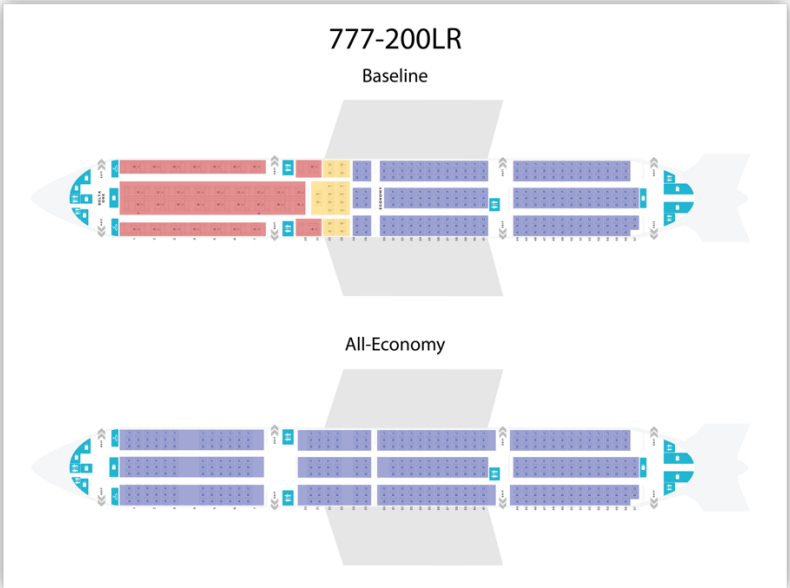
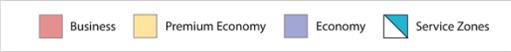

*Figure 1: Aircraft Cabin Space Allocation*

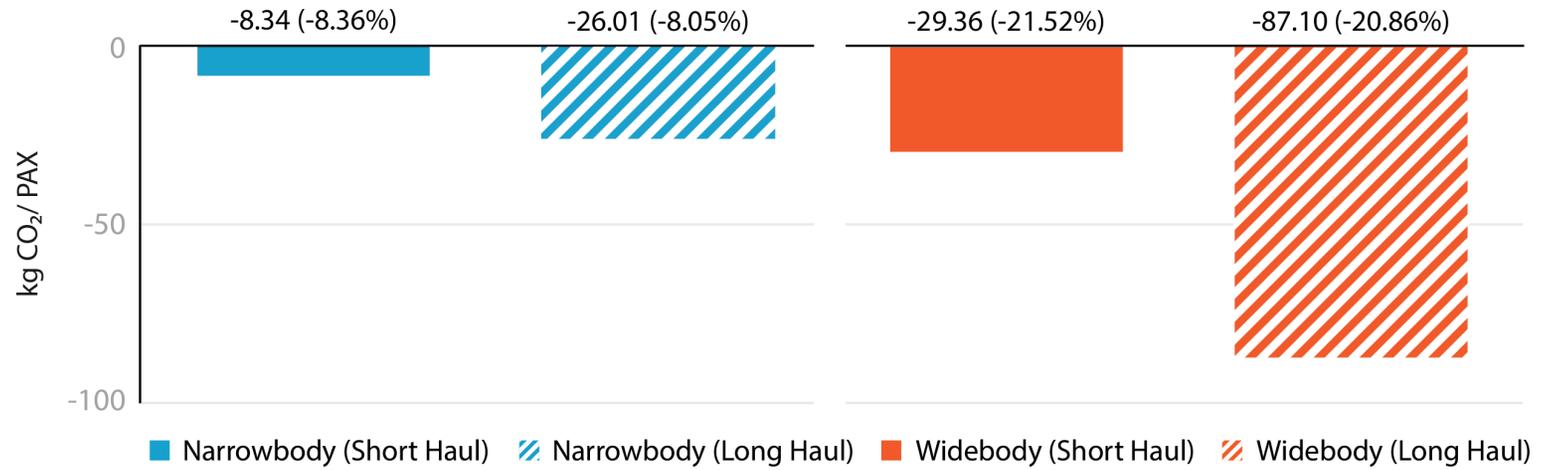

*Figure 2a: Absolute Economy-Seat Emissions Change from Baseline (kg $CO_2$/ PAX)*

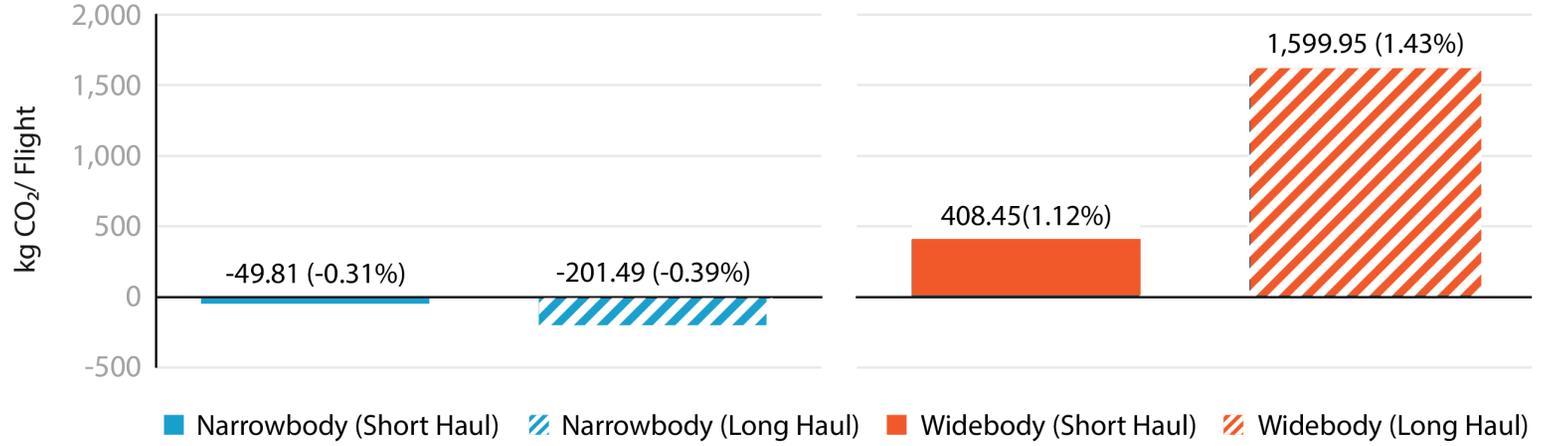

*Figure 2b: Absolute Emissions Change from Baseline (kg $CO_2$/ Flight)*

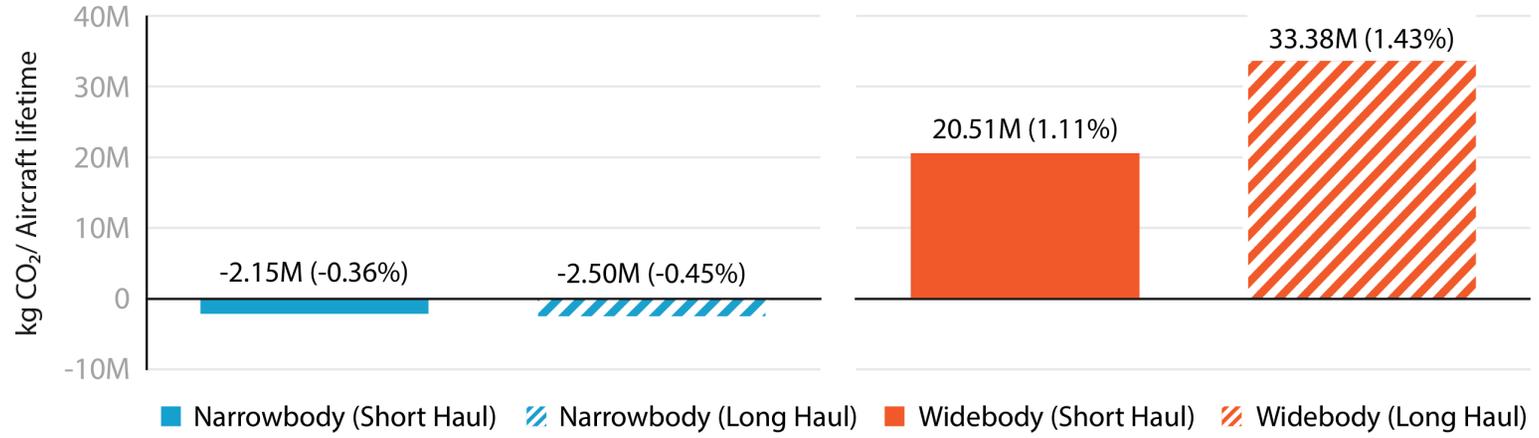

*Figure 2c: Absolute Emissions Change from Baseline (kg $CO_2$/ Aircraft lifespan)*

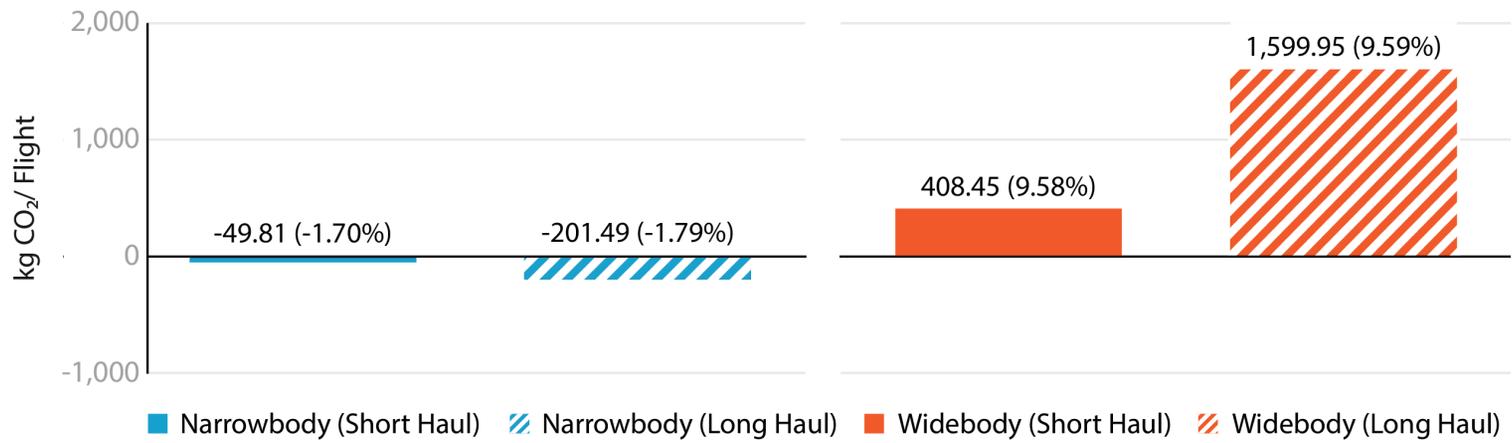

*Figure 3a: Absolute 'Variable' Emissions Change from Baseline (kg $CO_2$/ Flight)*

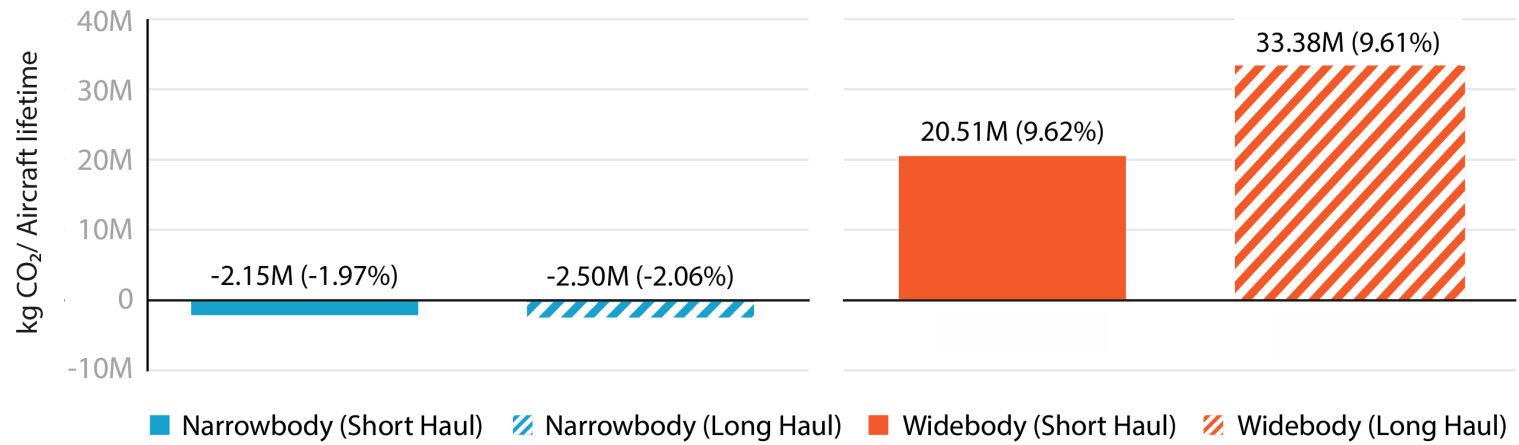

*Figure 3b: Absolute 'Variable' Emissions Change from Baseline (kg $CO_2$/ Aircraft lifespan)*

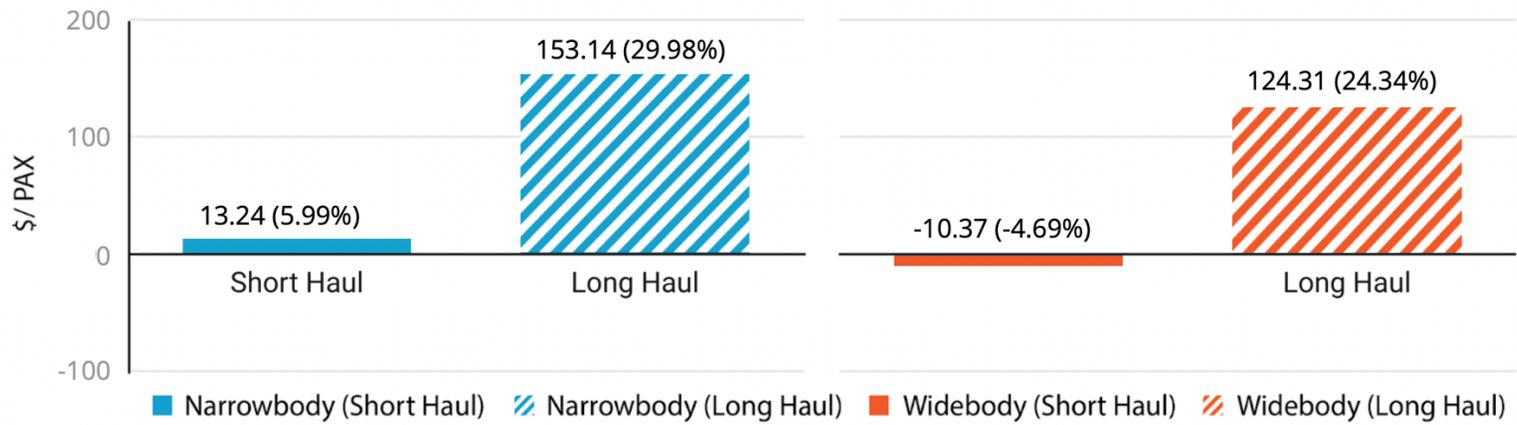

*Figure 4a: Absolute Economy-Seat Price Change from Baseline ($ / PAX)*

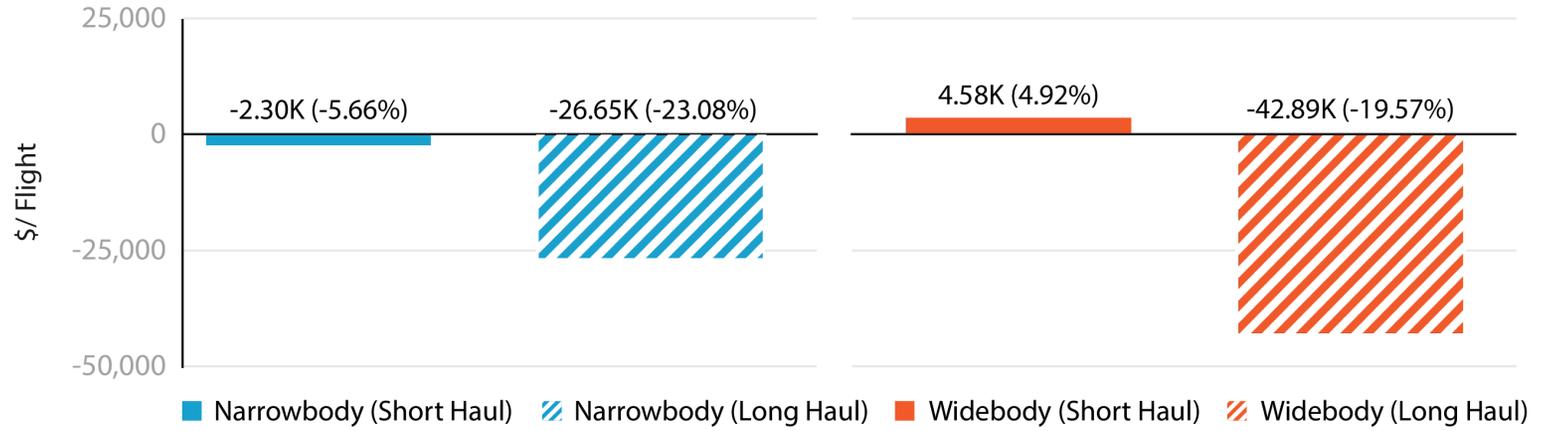

*Figure 4b: Absolute Revenue Change from Baseline ($ / Flight)*

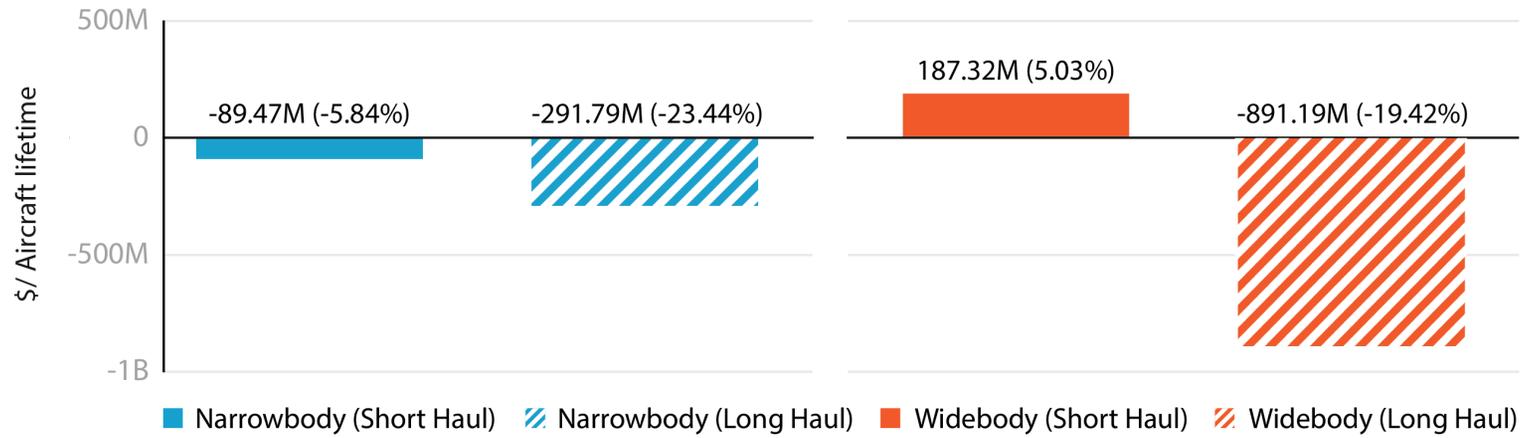

*Figure 4c: Absolute Revenue Change from Baseline ($ / Aircraft lifespan)*

## Supplementary Information

Here, we provide a summary of additional data and information that informs our model development. In Section A, we summarize the cabin allocation input parameters, including the configurations of the 16 airlines used to construct our four composite aircraft. Section B details the aircraft emissions input parameters, outlining our linear regression analyses and both the baseline emissions for an empty aircraft and the marginal emissions penalty per passenger. Section C describes the financial input parameters, including route-specific ticket prices across cabin classes, which collectively inform the financial aspect of our model. Finally, Section D provides a comprehensive summary of our results, reporting emissions and financial outcomes across different aircraft types, flight lengths, and scenarios.

*Section A: Cabin allocation input parameters*

In Table S1a, we provide a comprehensive summary of the cabin allocations for each of the 16 airlines used to construct our four composite aircraft, which we employ in our model.

Using the cabin allocations from 16 airlines – four airlines for each aircraft type – we compute the average of the four composite aircraft. Table S1b provides a summary of these configurations.

| | | Narrow-body | | | | | | | | Wide-body | | | | | | | |
| --- | --- | --- | --- | --- | --- | --- | --- | --- | --- | --- | --- | --- | --- | --- | --- | --- | --- |
| | Aircraft | A320-200 | | | | 737-800 | | | | A330-200 | | | | 777-200LR | | | |
| | Airline | Air Canada | Delta Airlines | ITA Airways | Turkish Airlines | Delta Airlines | Japan Airlines | Qantas Airways | Turkish Airlines | Air France | Delta Airlines | ITA Airways | Turkish Airlines | Air Canada | Delta Airlines | Ethiopian Airlines | Qatar Airways |
| Cabin Seating Quantities (#) | Economy | 132 | 123 | 128 | 141 | 108 | 132 | 162 | 135 | 147 | 168 | 219 | 190 | 236 | 220 | 287 | 230 |
| | Premium Economy | 0 | 18 | 37 | 0 | 36 | 0 | 0 | 0 | 21 | 32 | 19 | 0 | 24 | 48 | 0 | 0 |
| | Business | 14 | 16 | 0 | 12 | 16 | 12 | 12 | 20 | 40 | 34 | 20 | 30 | 40 | 28 | 28 | 42 |
| Cabin Allocation (%) | Economy | 46 | 46.6 | 50.4 | 62 | 38.84 | 49.4 | 55.7 | 53 | 31 | 34.06 | 39.9 | 33.56 | 32.2 | 34.4 | 52 | 33.3 |
| | Premium Economy | 0.00 | 9.42 | 16.50 | 0.00 | 14.30 | 0.00 | 0.00 | 0.00 | 7.90 | 7.28 | 5.00 | 0.00 | 5.20 | 11.70 | 0.00 | 0.00 |
| | Business | 15.90 | 13.00 | 0.00 | 9.00 | 17 | 11.60 | 7.80 | 10.30 | 20.50 | 19.20 | 13.30 | 21.00 | 24.80 | 17.00 | 10.00 | 20.00 |
| | Service Zones | 38.10 | 30.98 | 33.10 | 29.00 | 29.86 | 39.00 | 36.50 | 36.70 | 40.60 | 39.46 | 41.80 | 45.44 | 37.80 | 36.90 | 38.00 | 46.70 |
| Cabin Spatial Allocation (sq ft) | Economy | 2,215.25 | 2,244.15 | 2,427.15 | 2,985.78 | 2,064.31 | 2,625.56 | 2,960.40 | 2,816.89 | 3,413.30 | 3,750.23 | 4,393.25 | 3,695.17 | 4,113.88 | 4,394.95 | 6,643.53 | 4,254.41 |
| | Premium Economy | 0.00 | 453.65 | 794.60 | 0.00 | 760.03 | 0.00 | 0.00 | 0.00 | 869.84 | 801.58 | 550.53 | 0.00 | 664.35 | 1,494.79 | 0.00 | 0.00 |
| | Business | 765.71 | 626.05 | 0.00 | 433.42 | 903.53 | 616.53 | 414.56 | 547.43 | 2,257.18 | 2,114.04 | 14,64.42 | 2,312.24 | 3,168.45 | 2,171.92 | 1,277.60 | 2,555.20 |
| | Service Zones | 1,834.81 | 1,491.92 | 1,594.02 | 1,396.57 | 1,587.03 | 2,072.81 | 1,939.94 | 1,950.57 | 4,470.32 | 4,344.80 | 4,602.45 | 5,003.24 | 4,829.33 | 4,714.35 | 4,854.88 | 5,966.40 |
| Seating Spatial Allocation (sq ft) | Economy | 16.78 | 18.25 | 18.96 | 21.18 | 19.11 | 19.89 | 18.27 | 20.87 | 23.22 | 22.32 | 20.06 | 19.45 | 17.43 | 19.98 | 23.15 | 18.5 |
| | Premium Economy | - | 25.20 | 21.48 | - | 21.11 | - | - | - | 41.42 | 25.05 | 28.98 | - | 27.68 | 31.14 | - | - |
| | Business | 54.69 | 39.13 | - | 36.12 | 56.47 | 51.38 | 34.55 | 27.37 | 56.43 | 62.18 | 73.22 | 77.07 | 79.21 | 77.57 | 45.63 | 60.84 |
| | Service Zones | 12.57 | 9.50 | 9.66 | 9.13 | 9.92 | 14.39 | 11.15 | 12.58 | 21.49 | 18.57 | 17.84 | 22.74 | 16.10 | 15.93 | 15.41 | 21.94 |

*Table S1a: Space Distribution by Airline*

|  |  | Narrow-body | | Wide-body | |
| --- | --- | --- | --- | --- | --- |
| Aircraft | | A320-200 | 737-800 | A330-200 | 777-200LR |
| Cabin Seating Quantities (#) | Economy | 131 | 134.25 | 181 | 243.25 |
| | Premium Economy | 13.75 | 9.00 | 18.00 | 18.00 |
| | Business | 10.50 | 15.00 | 31.00 | 34.50 |
| Cabin Allocation (%) | Economy | 51 | 49 | 35 | 38 |
| | Premium Economy | 6 | 4 | 5 | 4 |
| | Business | 9 | 12 | 19 | 18 |
| | Service Zones | 33 | 36 | 42 | 40 |
| Cabin Spatial Allocation (sq ft) | Economy | 2,468.08 | 2,616.79 | 3,812.99 | 4,851.69 |
| | Premium Economy | 312.06 | 190.01 | 555.49 | 539.79 |
| | Business | 456.29 | 620.51 | 2,036.97 | 2,293.29 |
| | Service Zones | 1,579.33 | 1,887.59 | 4,605.20 | 5,091.24 |
| Seating Spatial Allocation (sq ft) | Economy | 18.79 | 19.54 | 21.26 | 19.76 |
| | Premium Economy | 23.34 | 21.11 | 31.82 | 29.41 |
| | Business | 43.31 | 42.44 | 67.23 | 65.81 |
| | Service Zones | 10.22 | 12.01 | 20.16 | 17.35 |

*Table S1b: Space Distribution by Composite Aircraft*

*Section B: Aircraft emission input parameters*

We source emissions data across 47 routes (24 short-haul and 23 long-haul), varying the load factor (i.e., the number of passengers on board) for each aircraft type. Then, we calculate the average emissions for each load factor, categorized by short-haul and long-haul flights, as well as by aircraft type. Table S2a-S2d provides a summary – linking aircraft emissions to the number of passengers onboard.

From here, we estimate the linear relationship between emissions and passenger count, allowing us to calculate the total aircraft emissions for any given number of passengers. Figure S1 visualizes these linear regressions, and Table S2e provides a summary.

Table S2f summarizes the emissions input parameters, including the baseline emissions for an empty aircraft and the emissions factor, which represents the incremental emissions per additional kilogram of onboard weight.

*Note: While the y-intercepts observed in Table S2e are implied as the emissions for carrying an aircraft with no passengers, those numbers differ from the 'empty aircraft emissions' listed in Table S2f because the linear regression assumes a standard seating configuration on board. In contrast, our 'empty aircraft emissions' exclude emissions for the seats. Our model accounts for the seat emissions later on, given the cabin class allocations. The process for adjusting the y-intercepts to remove seat emissions is detailed in the Methods section. We also detail the process for adjusting the slope (i.e., emissions for an additional passenger) into the 'emissions factor' (i.e., emissions for an additional kilogram of weight) in the Methods section.*

| Short Haul | | Long Haul | |
|---|---|---|---|
| Passengers (#) | Emissions (kg $CO_2$/flight) | Passengers (#) | Emissions (kg $CO_2$/flight) |
| 0 | 13,216.25 | 0 | 41,668.90 |
| 95 | 14,755.85 | 95 | 46,462.71 |
| 190 | 16,116.80 | 190 | 52,123.00 |

*Table S2a: Input Data for A320-200 Linear Regression of Passenger Count and Emissions*

| Short Haul | | Long Haul | |
|---|---|---|---|
| Passengers (#) | Emissions (kg $CO_2$/flight) | Passengers (#) | Emissions (kg $CO_2$/flight) |
| 0 | 14,142.55 | 0 | 44,641.86 |
| 95 | 15,244.20 | 95 | 49,906.75 |
| 189 | 16,764.80 | 189 | 55,435.53 |

*Table S2b: Input Data for 737-800 Linear Regression of Passenger Count and Emissions*

| Short Haul | | Long Haul | |
|---|---|---|---|
| Passengers (#) | Emissions (kg $CO_2$/flight) | Passengers (#) | Emissions (kg $CO_2$/flight) |
| 0 | 30,271.90 | 0 | 90,917.95 |
| 115 | 31,833.35 | 115 | 96,477.50 |
| 230 | 33,420.55 | 230 | 102,649.10 |

*Table S2c: Input Data for A330-200 Linear Regression of Passenger Count and Emissions*

| Short Haul | | Long Haul | |
|---|---|---|---|
| Passengers (#) | Emissions (kg $CO_2$/flight) | Passengers (#) | Emissions (kg $CO_2$/flight) |
| 0 | 108,191.75 | 0 | 36,123.70 |
| 210 | 116,957.00 | 210 | 38,230.05 |
| 300 | 121,338.55 | 300 | 39,345.35 |

*Table S2d: Input Data for 777-200LR Linear Regression of Passenger Count and Emissions*

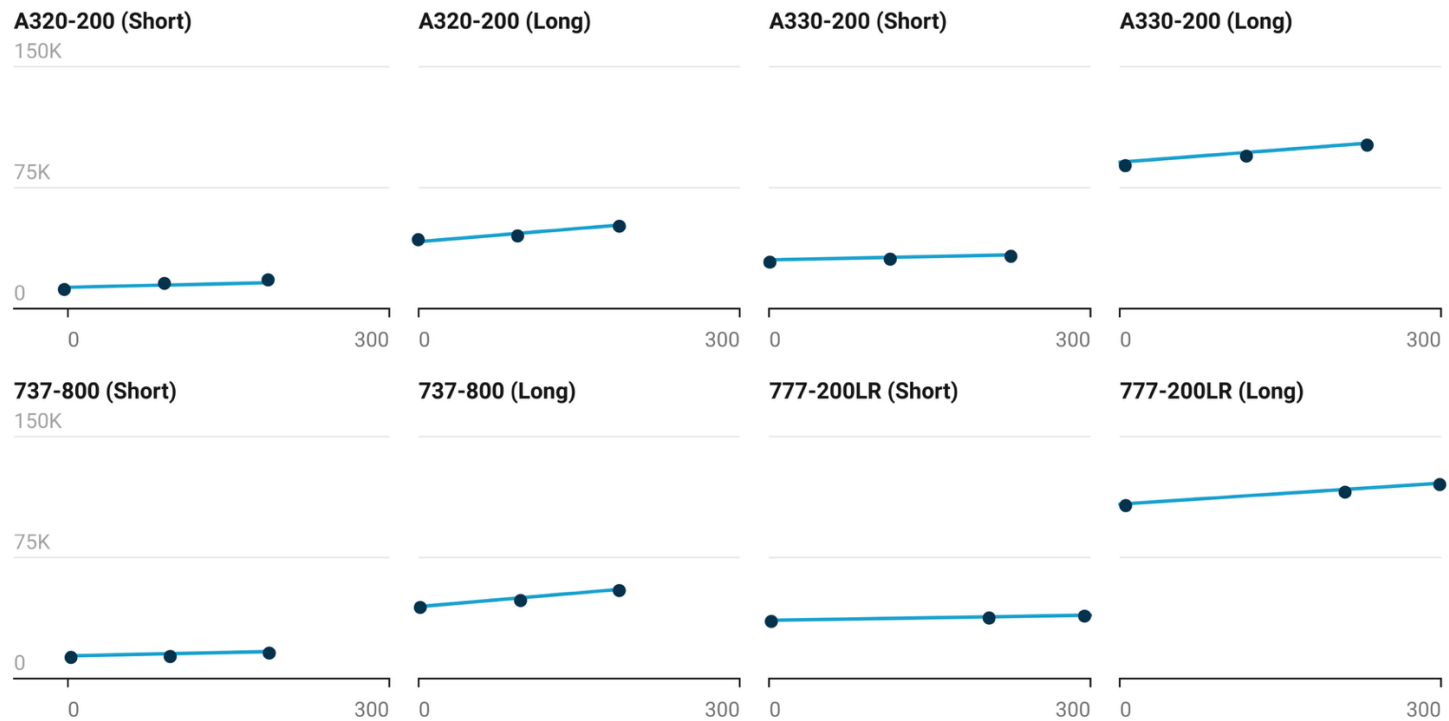

*Figure S1: Linear Regressions of Passenger Count and Emissions (*kg CO$_2$/flight)

| Aircraft | Stage Length | Intercept (kg $CO_2$) | Slope (kg $CO_2$/ PAX) |
|---|---|---|---|
| A320-200 | Short Haul | 13,246 | 15.30 |
| A320-200 | Long Haul | 41,524 | 55.00 |
| 737-800 | Short Haul | 14,073 | 13.90 |
| 737-800 | Long Haul | 44,589 | 57.10 |
| A330-200 | Short Haul | 30,268 | 13.70 |
| A330-200 | Long Haul | 90,816 | 51.00 |
| 777-200LR | Short Haul | 36,096 | 10.60 |
| 777-200LR | Long Haul | 108,112 | 43.50 |

*Table S2e: Linear Regression Results*

| Aircraft | Stage Length | Empty Aircraft Emissions (kg CO2) | Emissions Factor (kg CO2/ kg Weight) |
|---|---|---|---|
| A320-200 | Short Haul | 12,622.64 | 0.20 |
| A320-200 | Long Haul | 39,283.16 | 0.73 |
| 737-800 | Short Haul | 13,412.04 | 0.19 |
| 737-800 | Long Haul | 41,873.84 | 0.76 |
| A330-200 | Short Haul | 29,101.76 | 0.18 |
| A330-200 | Long Haul | 86,474.52 | 0.68 |
| 777-200LR | Short Haul | 35,007.68 | 0.14 |
| 777-200LR | Long Haul | 103,645.76 | 0.58 |

*Table S2f: Emissions Breakdown*

*Section C: Financial input parameters*

Table S3a and S3b report the 90-day rolling average ticket prices for economy, premium economy, and business class seats across 24 airlines, for each of the 24 short-haul and 23 long-haul routes, respectively.

Table S3c provides a summary of our financial data by averaging the 90-day rolling ticket prices across all 24 airlines for each short-haul and long-haul route. For each haul type, we calculate a single average price for economy, premium economy, and business class, yielding six aggregate values that are used in our model and displayed as inputs.

| Airline | Route | Economy | Premium Economy | Business |
|---|---|---|---|---|
| Delta Airlines | ORD - JFK | 169.65 | 217.70 | 297.25 |
| Delta Airlines | ATL - BOS | 317.31 | 404.28 | 792.15 |
| American Airlines | LAX - SEA | 191.98 | 263.84 | 260.49 |
| American Airlines | MIA-DCA | 163.10 | 242.10 | 745.66 |
| United Airlines | ORD - EWR | 153.90 | 233.82 | 311.76 |
| United Airlines | MSO - DEN | 239.33 | 313.25 | 343.61 |
| Aeroméxico | MEX-MIA | 309.69 | 392.92 | 579.83 |
| Aeroméxico | MEX-CUN | 171.48 | 184.49 | 197.54 |
| Emirates | DXB-BOM | 164.67 | 546.67 | 832.45 |
| Emirates | DXB-BAH | 213.54 | 535.21 | 1,335.26 |
| China Southern | PKX-SHA | 233.16 | 256.56 | 634.92 |
| China Southern | PKX-HGH | 339.88 | 374.22 | 935.22 |
| EVA | TPE-SGN | 293.89 | 453.10 | 646.47 |
| EVA | TPE-BKK | 309.30 | 443.02 | 664.89 |
| China Airlines | TPE-PVG | 292.64 | 471.71 | 675.51 |
| China Airlines | TPE-SGN | 177.37 | 449.28 | 637.93 |
| Air India | DEL-BLR | 95.70 | 123.11 | 412.31 |
| Air India | DEL-BOM | 102.06 | 121.19 | 385.37 |
| Air New Zealand | SYD-AKL | 231.11 | 431.31 | 671.91 |
| Air New Zealand | BNE-AKL | 250.61 | 420.61 | 661.55 |

*Table S3a: 90-Day Rolling Average of Short-Haul Flights ($)*

| Airline | Route | Economy | Premium Economy | Business |
|---|---|---|---|---|
| Delta Airlines | JFK-LHR | 423.53 | 841.00 | 2,179.91 |
| Delta Airlines | BOS-CDG | 494.74 | 1,286.24 | 2,218.98 |
| American Airlines | PHL-LHR | 494.88 | 1,537.17 | 3,710.13 |
| American Airlines | IAD-LHR | 420.72 | 1,282.12 | 2,526.55 |
| United Airlines | EWR-LHR | 331.96 | 1,037.93 | 2,068.85 |
| United Airlines | EWR-CDG | 368.46 | 1,087.93 | 5,307.33 |
| Aeroméxico | MEX-SEA | 546.04 | 608.04 | 911.55 |
| Aeroméxico | LIM-MEX | 557.10 | 643.47 | 790.21 |
| ANA | KUL-NRT | 629.62 | 1,002.17 | 1,845.26 |
| ANA | SIN-NRT | 916.22 | 1,054.73 | 1,716.87 |
| Air Canada | YYZ-LHR | 686.78 | 1,202.08 | 7,431.70 |
| Air Canada | YYZ-CDG | 702.10 | 1,209.70 | 5,548.73 |
| Emirates | DXB-GVA | 639.08 | 1,176.53 | 3,177.54 |
| Emirates | DXB-LHR | 553.54 | 1,102.79 | 3,109.42 |
| Singapore Airlines | SIN-NRT | 685.14 | 1,381.69 | 2,401.76 |
| Singapore Airlines | SIN-BOM | 272.33 | 905.20 | 1,114.66 |
| Air France | JFK-CDG | 376.49 | 1,193.24 | 2,395.64 |
| Air France | BOS-CDG | 417.15 | 1,375.93 | 2,551.82 |
| British Airways | JFK-LHR | 356.67 | 1,638.87 | 2,152.38 |
| British Airways | BOS-LHR | 343.39 | 1,683.87 | 1,990.58 |

*Table S3b: 90-Day Rolling Average of Long-Haul Flights ($)*

|                 | Short Haul | Long Haul |
|-----------------|------------|-----------|
| Economy         | 221.02     | 510.80    |
| Premium Economy | 343.92     | 1,162.53  |
| Business        | 601.10     | 2,757.49  |

*Table S3c: Financial Dataset Summary ($)*

*Section D: Results*

Below, we provide figures (Fig. S2a-S3) that summarize the total emissions on a per-passenger, per-flight, and over the aircraft's lifespan basis, as well as total revenue per flight.

Additionally, we provide tables (Table S4ai-S4ciii) summarizing the results of our model, detailing emissions from air travel under both the baseline and revised scenarios. We break down results by aircraft type (narrow-body and wide-body), flight length (short-haul and long-haul), and report results per-passenger, per-flight, and over the aircraft's lifespan. We show aggregate emissions and both the absolute and relative change in emissions when transitioning from the baseline to an all-economy configuration.

We then apply the same methodology to analyze only 'variable' emissions (Table S4di-S4fii). For financial outcomes, we provide analogous results for ticket prices and airline revenue (Table S5ai-S5cii).

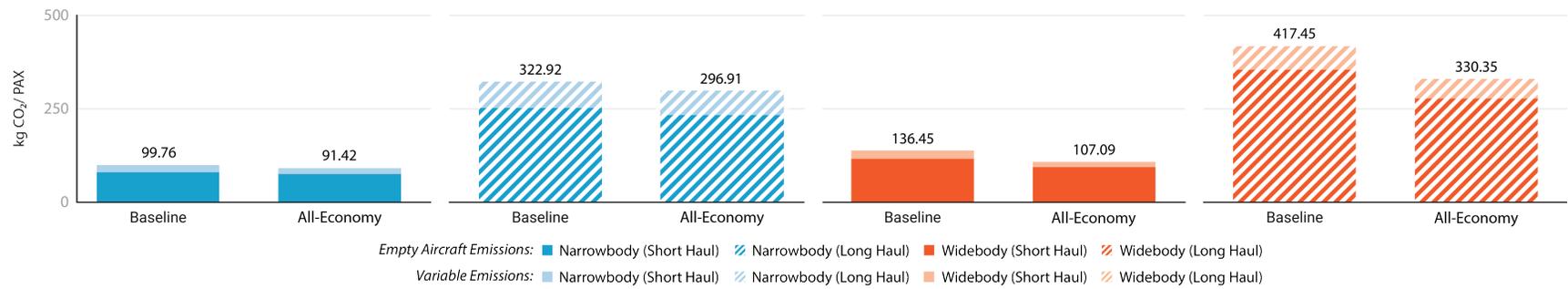

*Figure S2a: Total Economy-Seat Emissions (kg $CO_2$ / PAX)*

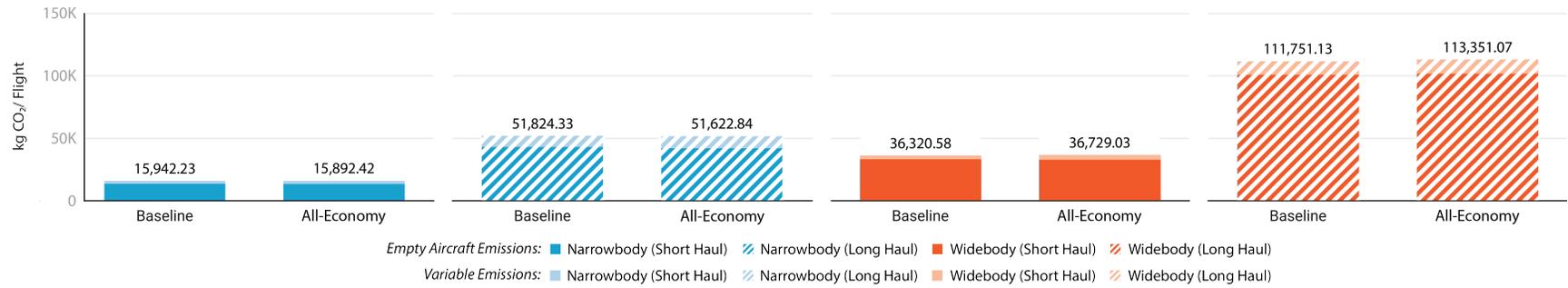

*Figure S2b: Total Emissions (kg CO$_2$/ Flight)*

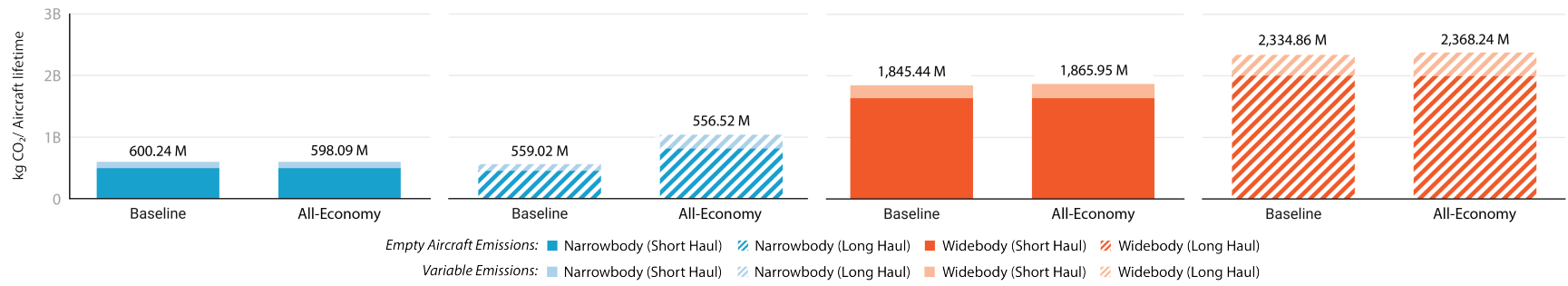

*Figure S2c: Total Emissions (kg $CO_2$/ Aircraft lifespan)*

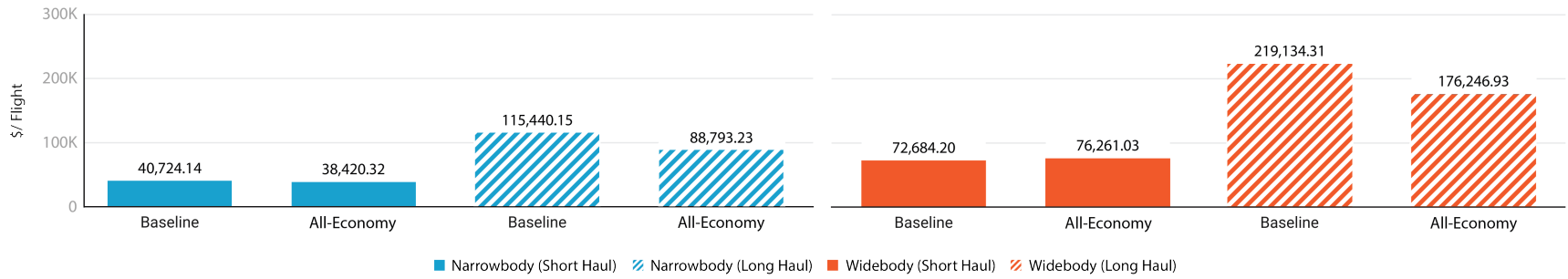

*Figure S3: Total Revenue ($ / Flight)*

| Body Type | Scenario | Short Haul | Long Haul |
|---|---|---|---|
| Narrow-body | Baseline | 99.76 | 322.92 |
| Narrow-body | All-Economy | 91.42 | 296.91 |
| Wide-body | Baseline | 136.45 | 417.45 |
| Wide-body | All-Economy | 107.09 | 330.35 |

*Figure S4ai: Total Economy-Seat Emissions (kg $CO_2$ / PAX)*

| Body Type | Short Haul | Long Haul |
|---|---|---|
| Narrow-body | -8.36 | -8.05 |
| Wide-body | -21.52 | -20.86 |

*Figure S4aii: Relative Economy-Seat Emissions Change from Baseline (%Δ / PAX)*

| Body Type | Short Haul | Long Haul |
|---|---|---|
| Narrow-body | -8.34 | -26.01 |
| Wide-body | -29.36 | -87.1 |

*Figure S4aiii: Absolute Economy-Seat Emissions Change from Baseline (kg $CO_2$/ PAX)*

| Body Type | Scenario | Short Haul | Long Haul |
|---|---|---|---|
| Narrow-body | Baseline | 15,942.23 | 51,824.33 |
| Narrow-body | All-Economy | 15,892.42 | 51,622.84 |
| Wide-body | Baseline | 36,320.58 | 111,751.13 |
| Wide-body | All-Economy | 36,729.03 | 113,351.07 |

Table S4bi: Total Emissions (kg $CO_2$/ Flight)

| Body Type | Short Haul | Long Haul |
|---|---|---|
| Narrow-body | -0.31 | -0.39 |
| Wide-body | 1.12 | 1.43 |

Table S4bii: Relative Emissions Change from Baseline (%Δ / Flight)

| Body Type | Short Haul | Long Haul |
|---|---|---|
| Narrow-body | -49.81 | -201.49 |
| Wide-body | 408.45 | 1,599.95 |

Table S4biii: Absolute Emissions change from Baseline (kg $CO_2$/ Flight)

| Body Type | Scenario | Short Haul | Long Haul |
|---|---|---|---|
| Narrow-body | Baseline | 600,240,713.63 | 559,021,223.85 |
| Narrow-body | All-Economy | 598,090,519.47 | 556,524,351.10 |
| Wide-body | Baseline | 1,845,443,808.50 | 2,334,857,788.84 |
| Wide-body | All-Economy | 1,865,950,020.49 | 2,368,236,048.77 |

*Table S4ci: Total Emissions (kg $CO_2$/ Aircraft lifespan)*

| Body Type | Short Haul | Long Haul |
|---|---|---|
| Narrow-body | -0.36 | -0.45 |
| Wide-body | 1.11 | 1.43 |

*Table S4cii: Relative Emissions Change from Baseline (%Δ / Aircraft lifespan)*

| Body Type | Short Haul | Long Haul |
|---|---|---|
| Narrow-body | -2,150,194.15 | -2,496,872.75 |
| Wide-body | 20,506,212.00 | 33,378,259.92 |

*Table S4ciii: Absolute Emissions Change from Baseline (kg $CO_2$/ Aircraft lifespan)*

| Body Type | Short Haul | Long Haul |
|---|---|---|
| Narrow-body | -16.75 | -16.70 |
| Wide-body | -11.54 | -11.63 |

*Figure S4di: Relative 'Variable' Economy-Seat Emissions Change from Baseline (%Δ / PAX)*

| Body Type | Short Haul | Long Haul |
|---|---|---|
| Narrow-body | -2.77 | -10.74 |
| Wide-body | -2.16 | -8.36 |

*Figure S4dii: Absolute 'Variable' Economy-Seat Emissions Change from Baseline (kg $CO_2$/ PAX)*

| Body Type | Short Haul | Long Haul |
|---|---|---|
| Narrow-body | -1.70 | -1.79 |
| Wide-body | 9.58 | 9.59 |

Table S4ei: Relative 'Variable' Emissions Change from Baseline (%Δ / Flight)

| Body Type | Short Haul | Long Haul |
|---|---|---|
| Narrow-body | -49.81 | -201.49 |
| Wide-body | 408.45 | 1,599.95 |

Table S4eii: Absolute 'Variable' Emissions Change from Baseline (kg $CO_2$/ Flight)

| Body Type | Short Haul | Long Haul |
|---|---|---|
| Narrow-body | -1.97 | -2.06 |
| Wide-body | 9.62 | 9.61 |

Table S4fi: Relative 'Variable' Emissions Change from Baseline (%Δ / Aircraft lifespan)

| Body Type | Short Haul | Long Haul |
|---|---|---|
| Narrow-body | -2,150,194 | -2,496,873 |
| Wide-body | 20,506,212 | 33,378,260 |

Table S4fii: Absolute 'Variable' Emissions Change from Baseline (kg $CO_2$/ Aircraft lifespan)

| Body Type | Short Haul | Long Haul |
|---|---|---|
| Narrow-body | 5.99 | 29.98 |
| Wide-body | -4.69 | 24.34 |

*Table S5ai: Relative Economy-Seat Price Change from Baseline (%Δ / PAX)*

| Body Type | Short Haul | Long Haul |
|---|---|---|
| Narrow-body | 13.24 | 153.14 |
| Wide-body | -10.37 | 124.31 |

*Table S5aii: Absolute Economy-Seat Price Change from Baseline ($ / PAX)*

| Body Type | Scenario | Short Haul | Long Haul |
|---|---|---|---|
| Narrow-body | Baseline | 40,724.14 | 115,440.15 |
| Narrow-body | All-Economy | 38,420.32 | 88,793.23 |
| Wide-body | Baseline | 72,684.20 | 219,134.31 |
| Wide-body | All-Economy | 76,261.03 | 176,246.93 |

*Table S5bi: Total Revenue ($ / Flight)*

| Body Type | Short Haul | Long Haul |
|---|---|---|
| Narrow-body | -5.66 | -23.08 |
| Wide-body | 4.92 | -19.57 |

*Table S5bii: Relative Revenue Change from Baseline (%Δ / Flight)*

| Body Type | Short Haul | Long Haul |
|---|---|---|
| Narrow-body | -2,303.82 | -26,646.92 |
| Wide-body | 3,576.83 | -42,887.38 |

*Table S5biii: Absolute Revenue Change from Baseline ($ / Flight)*

| Body Type | Short Haul | Long Haul |
|---|---|---|
| Narrow-body | -5.84 | -23.44 |
| Wide-body | 5.03 | -19.42 |

*Table S5ci: Relative Revenue Change from Baseline (%Δ / Aircraft lifespan)*

| Body Type | Short Haul | Long Haul |
|---|---|---|
| Narrow-body | -89,469,309.36 | -291,794,325.15 |
| Wide-body | 187,318,768.49 | -891,187,320.57 |

*Table S5cii: Absolute Revenue Change from Baseline ($ / Aircraft lifespan*